\documentclass[11pt]{article}
\usepackage{moriond,epsfig}

\def\be{\begin{equation}}
\def\ee{\end{equation}}
\def\bea{\begin{eqnarray}}
\def\eea{\end{eqnarray}}

\def\lsim{\raise0.3ex\hbox{$<$\kern-0.75em\raise-1.1ex\hbox{$\sim$}}}
\def\gsim{\raise0.3ex\hbox{$>$\kern-0.75em\raise-1.1ex\hbox{$\sim$}}}

\def\beq{\begin{equation}}
\def\eeq{\end{equation}}
\def\bea{\begin{eqnarray}}
\def\eea{\end{eqnarray}}
\def\bq{\begin{quote}}
\def\eq{\end{quote}}
\begin{document}
\vspace*{4cm}
\title{ELLIPTIC FLOW IN A FINAL STATE INTERACTION MODEL}

\author{A. CAPELLA$^{(a)}$,
\underline{E. G. FERREIRO}$^{(b)}$}

\address{$^{(a)}$ Laboratoire de Physique Th\'eorique, Universit\'e de Paris
XI, 
F-91405 Orsay Cedex, France\\
$^{(b)}$ Depto. de F\'{\i}sica de Part\'{\i}culas,
Universidade de Santiago de Compostela \\
E-15782 Santiago de Compostela, Spain}

\maketitle\abstracts{
We propose a final state interaction model
to describe the fixed $p_T$
suppression of the yield of particles at all values of $p_T$.
We make an extension of the model to the motion in the
transverse plane which introduces a dependence of the suppression on
the azimuthal angle $\theta_R$. We
obtain values of the elliptic flow $v_{2}$ close to the experimental
ones for all values of $p_T$.}

\section{Introduction}
The RHIC data show azimuthal anisotropy in the production of particles in heavy ion collsions.
This asymmetry is currently named elliptic flow $v_2$.
Two main interpretations has been applied in order to explain the data.
In the framework of the hydrodynamical models, it is considered as a
signal of early thermalization (Quark Gluon Plasma formation):
particles tend to go in the direction of the strongest pressure gradients,
hence preferably in the collision plan.
On the other hand, at larger transverse momenta, measurements of azimuthal
anisotropy are also relevant to the observation of jet quenching -- final state interactions.
In the second scenario, due to the asymmetry of the overlap region of the two nuclei,
the average amount of matter traversed by a parton depends on its azimuthal direction
with respect to the reaction plane, which leads to an azimuthal anisotropy of the emitted jets.
Our mechanism consists on a final state interactions approach applied to
the whole $p_T$ region. In this approach, we can reproduced the large $p_T$
suppression and the elliptic flow data.
In this way, the $v_2$ in our approach becomes medium dependent, while in other approaches
the $v_2$ corresponds to the gradient of the medium. 

\section{The final state interaction model}
The interaction of a particle or a parton with
the 
medium is
described by the gain and loss differential equations which govern
final state interactions:
\beq
\label{1e}
\tau\ {d\rho_i \over d\tau} = 
\sum_{k,\ell} \sigma_{k\ell} \ \rho_k\ \rho_{\ell} - \sum_k \sigma_{ik} \ \rho_i\ \rho_k \ ,
\eeq
where 
$\rho_i \equiv dN^{AA\to i}(b)/dy d^2s$ are transverse
densities and $\sigma_{ij}$ are the final state interaction
cross-sections.
The first term of (\ref{1e}) describes the gain in type $i$ particle
yield resulting from the interaction of $k$ and $\ell$. The second one
corresponds to the loss of type $i$ particles resulting from its
interaction with particle $k$.
Consider a $\pi^0$ produced at fixed $p_T$ interacting with
the hot medium.
In
the interaction, with cross-section $\sigma$, 
the $\pi^0$ suffers a decrease in its transverse momentum with a
$p_T$-shift $\delta p_T$. This produces a loss in the $\pi^0$ yield
in a given $p_T$ bin. There is also a gain resulting from $\pi^0$'s
produced at $p_T + \delta p_T$ and that are shifted to smaller values of $p_T$.
The gain and loss diferential equation for pions is then
\beq
\label{2e}
\tau {d\rho_{\pi^0} \over d\tau} = - \sigma \
\rho_{medium} \ \rho_{\pi^0}(b,s,y,p_T)
+ \sigma \
\rho_{medium} \ \rho_{\pi^0}(b,s,y,p_T+\delta p_T) \ .
\eeq
Due to the steep fall-off of the $p_T$
spectrum the loss is larger than the gain, resulting in a net
suppression of the $\pi^0$ yield at a given $p_T$.
Our equations have to be integrated
between initial time $\tau_0$ and freeze-out time $\tau_f$.
The
solution depends only on the ratio $\tau_f/\tau_0$.
We use the inverse proportionality between proper time and densities,
$\tau_f/\tau_0 = \rho(b,s,y)/\rho_{pp}(y)$, where 
$\rho_{pp}(y)$ corresponds to the density
per unit rapidity for 
$pp$ collisions at $\sqrt{s} = 200$
GeV= 2.24 fm$^{-2}$ and
$\rho(b,s,y)$ is the density produced in the primary
collisions.
Our densities
 can be either hadrons or partons. In fact,
at early
times, densities are very high and hadrons not yet formed, so 
our equations describe final state interactions at a
partonic level.
At later times we have interactions of full
fledged hadrons, and, thus,  
$\sigma$ represents an effective cross-section
averaged over the interaction time.

Integrating eq. (\ref{2e})
from $\tau_0$ to $\tau_f$ and taking
$\rho(b,y,p_T)_{\pi^0}=dN_{\pi^0}/db dy dp_T$,
we obtain
the suppression factor $S_{\pi^0}(b,y,p_T)$ of the yield of
$\pi^0$'s at given $p_T$ and at each impact parameter,
due to its
interaction with the dense medium:
\beq
\label{3e}
S_{\pi^0}(y, p_T, b) =
\exp \left \{ - \sigma \ \rho (b,y) \left [ 1 - {N_{\pi^0} (p_T + \delta p_T)
\over N_{\pi^0}(p_T)} (b)\right ] \
\ell n \left ({\rho (b,y) \over \rho_{pp}(y)} \right ) \right \} \ .
\eeq
When $\delta p_T$ tends to $\infty$,
the gain term vanishes, and  
the survival probability has the same expression as in
the case of $J/\psi$ suppression without $c\overline{c}$ recombination 
\cite{1r}.
If $\delta p_T$ is equal to 0, the loss and gain terms are identical and the
survival probability becomes one.

\section{Numerical results}
In order
to perform numerical
calculations, we need
the $p_T$ distribution of the $\pi^0$'s.
We use the following parametrization for 
the ratio $R_{AA}^0(b, p_T)$ in the absence of final
state interactions:
\beq
\label{4e}
\left . R_{AA}^0(b, p_T) = R_{AA}^0(b, p_T=0)
\left ( {p_T + p_0^{AA}(b) \over p_T + p_0^{pp}}\right )^{-n}\right / 
\left ( {p_0^{AA}(b) \over p_0^{pp}}\right )^{-n}
\nonumber
\eeq
where $p_0(b) = (n-3)/2<p_T>_b$, $n = 9.99$ and $<p_T>_b$ is the
experimental value of $<p_T>$ at each $b$. 
 We have also tried different parametrizations \cite{2r} for $R_{AA}^0(b, p_T)$.
Our final result
depends little on the form of $R_{AA}^0$ taking a $p_T$-shift $\delta p_T$ of the form
$p_T^{\alpha}/C$.
To the value $R^0_{AA}$ we apply the correction due to the suppression factor
$S_{\pi^0}$, $R_{AA}(b, p_T)=R_{AA}^0(b, p_T)\ S_{\pi^0}(b,p_T)$. 
Our results are shown in Fig. 1.
The dashed lines correponds to the result obtained using eq. ({\protect\ref{4e}}) for $R_{AuAu}^0$
and a $p_T$-shift given by $\delta p_T = {p_T^{3/2} / (20  \ {\rm GeV}^{1/2})}$.
The continous lines are obtain with $R_{AA}^0 (b,p_T \geq 5$~GeV) = 1  and the $p_T$-shift
given by $\delta p_T = {p_T^{3/2} / (20  \ {\rm GeV}^{1/2})}$ for $p_T < 2.9$~GeV
and $\delta p_T = {p_T^{0.8} / (9.5  \ {\rm GeV}^{1/2})}$
for $p_T \geq 2.9$~GeV.
\vskip -0.5cm
\begin{flushleft}
\begin{tabular}{cc}
\psfig{figure=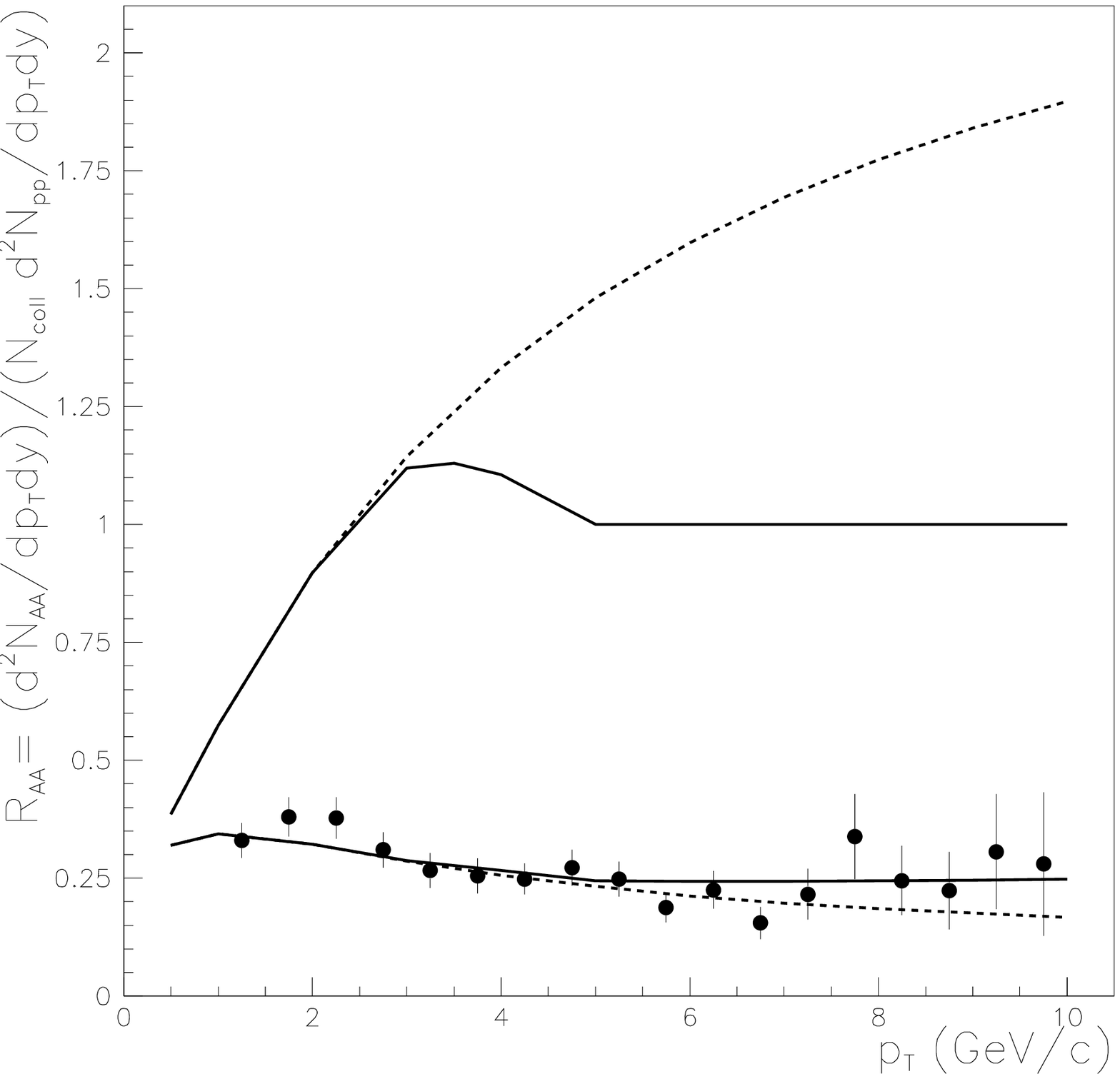,height=2.7in}
&
\psfig{figure=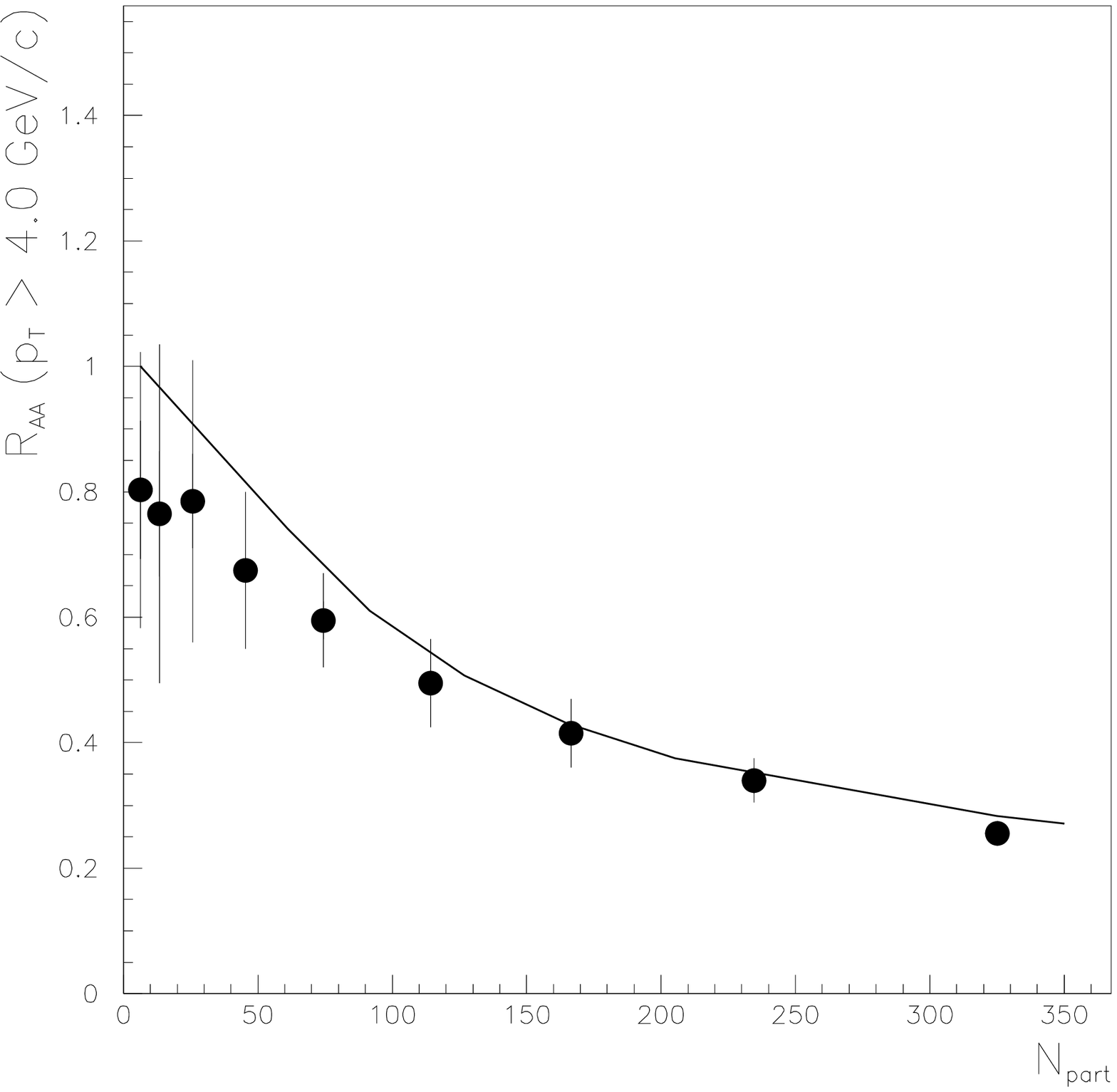,height=2.7in}\\
\end{tabular}
\end{flushleft}
\vskip -0.3cm
{\small Figure 1. Left: 
The $\pi^0$ suppression factor $R_{AuAu}(p_T,b)$ at $\sqrt{s} = 200$~GeV 
in the centrality bin 0-10 \% and the corresponding 
$R_{AuAu}^0(p_T,b)$
in the absence of final state interaction.\\
Right: Centrality dependence of $R_{AuAu}^{\pi^0}$
for $p_T > 4$~GeV/c. 
The data \cite{3r} are from PHENIX.}

\section{Elliptic flow}
Our final state interaction model 
takes into account the longitudinal expansion with no
consideration for the motion in the transverse plane.
Elliptic flow, on
the contrary, results from an asymmetry in the azimuthal angle, so
the motion in the transverse plane plays a fundamental role.
We propose a simple extension of the model \cite{4r} taking into account the different
path length of the particles or partons in the transverse plane for each value of
its azimuthal angle $\theta_R$ -- measured with respect to the reaction
plane.
At $y^* \sim 0$, the path length $R_{\theta_R}$, 
measured from the center of the interaction region (overlap of the colliding nuclei) 
is given by
\beq
\label{5e}
R_{\theta_R}(b) = R_A {\sin (\theta_R - \alpha) \over \sin \theta_R} \ ,
\eeq
where $R_A = 1.05 \ A^{1/3}$~fm is the nuclear radius and 
$\sin \alpha = b \sin  \theta_R/2R_A$. 
Our ansatz consists in the following replacement in eq. (\ref{3e})
\beq
\label{6e}
\rho (b, y) \to \rho (b, y) {R_{\theta_R} \over <R_{\theta_R}>}\ .
\eeq
This is motivated by the
fact that the duration of the interaction,
as well as the density of
the medium traversed by the particles, are expected to be proportional to
the path length $R_{\theta_R}$
inside the overlap region of the colliding nuclei.
Due to the division by $<R_{\theta_R}>$ in 
eq. (\ref{6e}), the results of section 3 are unchanged.
With the replacement (\ref{6e}), the survival probability becomes
$\theta_R$-dependent, 
and the elliptic flow can be obtained as
\beq
\label{7e}
v_{2}(p_T, b) = {\int_0^{90^{\circ}} d \theta_R \ \cos 2 \theta_R\ S_{\pi^0}^{\theta_R} (p_T, b) \over \int_0^{90^{\circ}} d\theta_R\ S_{\pi^0}^{\theta_R}(p_T, b)}\ .
\eeq
Clearly, when the particle or parton moves along the reaction plane $\theta_R = 0^{\circ}$
its path
length will have its minimal value and the survival probability its
maximal one. On the contrary, for $\theta_R = 90^{\circ}$ the path
length will be maximal and the survival probability minimal. 

Using the same value $\sigma = 1.3$~mb of the final state interaction
cross-section and the same $p_T$-shift introduced in section 3 
in order to describe
the experimental values of $R_{AA}(b, p_T)$, 
we obtain the values of
$v_{2}$ versus $p_T$ and centrality shown in Fig. 2.
\vskip -0.5cm
\begin{flushleft}
\begin{tabular}{cc}
\psfig{figure=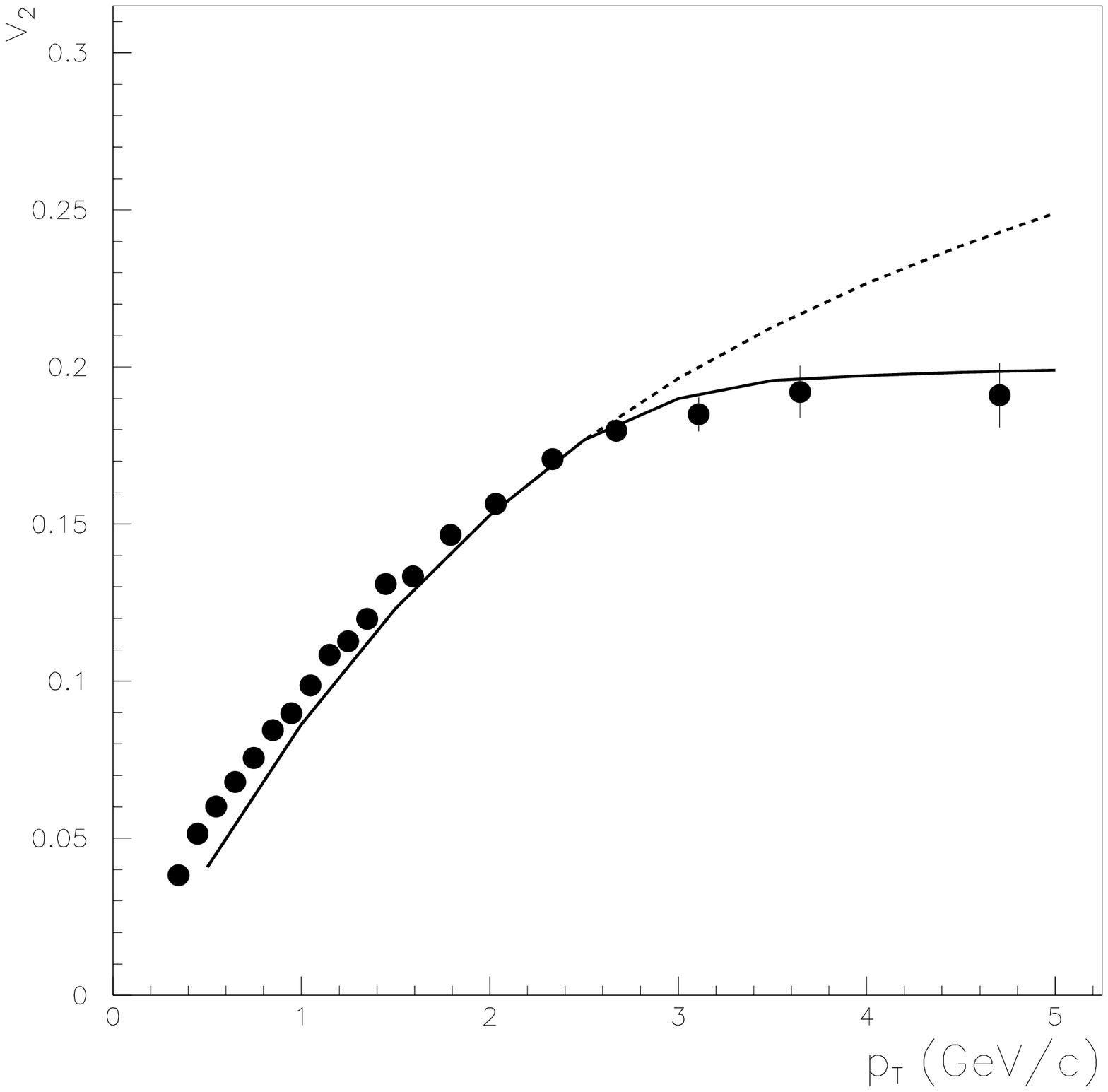,height=2.7in}
&
\psfig{figure=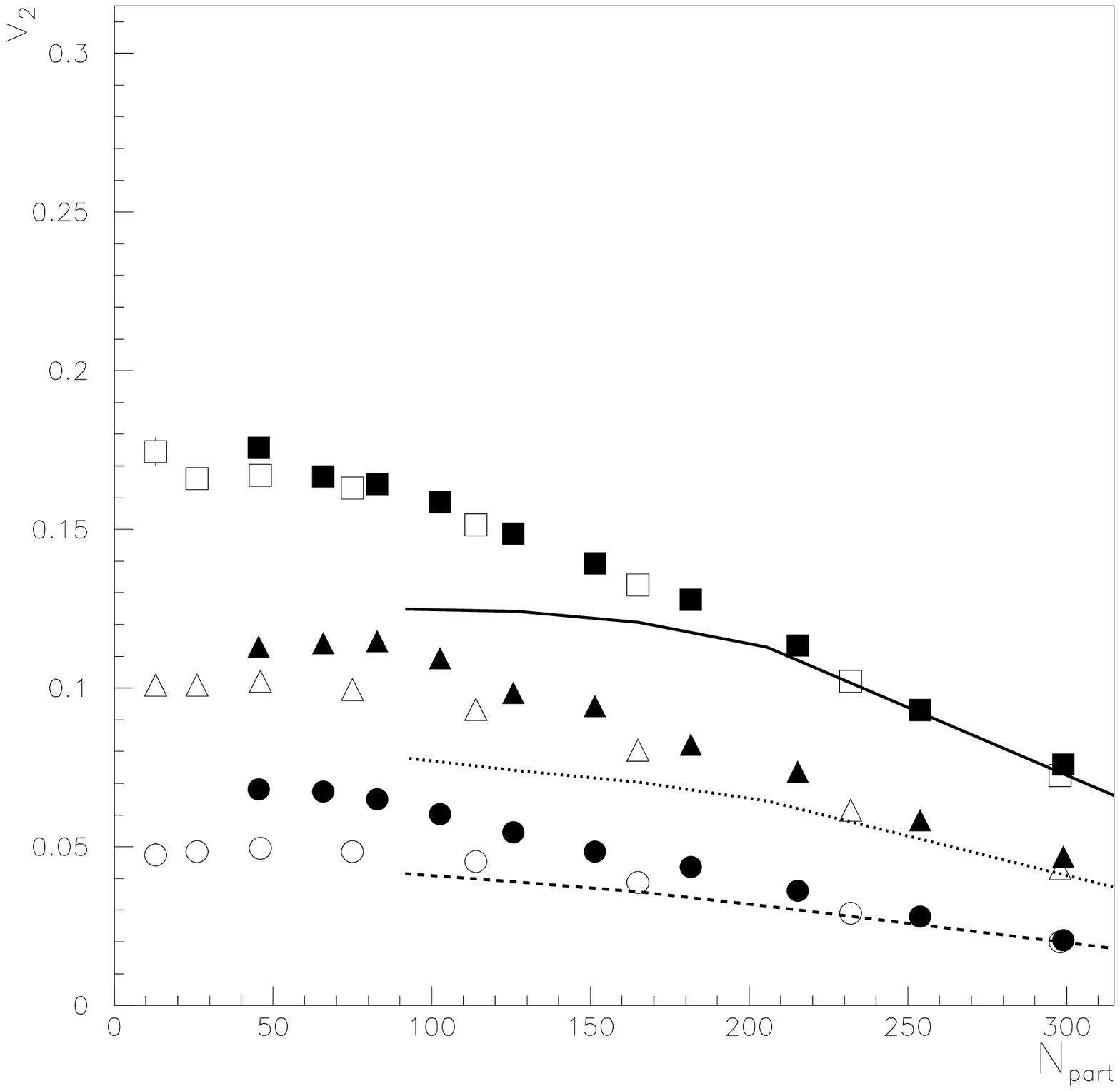,height=2.7in}\\
\end{tabular}
\end{flushleft}
\vskip -0.3cm
{\small Figure 2. Left:
$v_2$ vs. $p_T$
at centrality 13 \%-26~\%.
Right: $v_2$ vs. the number of participants 
for different values
of $p_T$: 0.4 (lower), 0.75 (middle) and 1.35 GeV (top).
Data \cite{3r}: PHENIX (black), STAR (open).}

\noindent
We find a good agreement with data for the $p_T$ and the centrality dependence.
For the mass dependence, in Fig. 3, 
our $v_2$ of mesons falls below that of baryons for $p_T > 2$ GeV, in agreement
with 
data, while the 
hydrodynamical model predicts the same mass ordering for $v_2$ at all $p_T$.
\vskip -2.0cm
\begin{flushleft}
\begin{tabular}{cc}
\psfig{figure=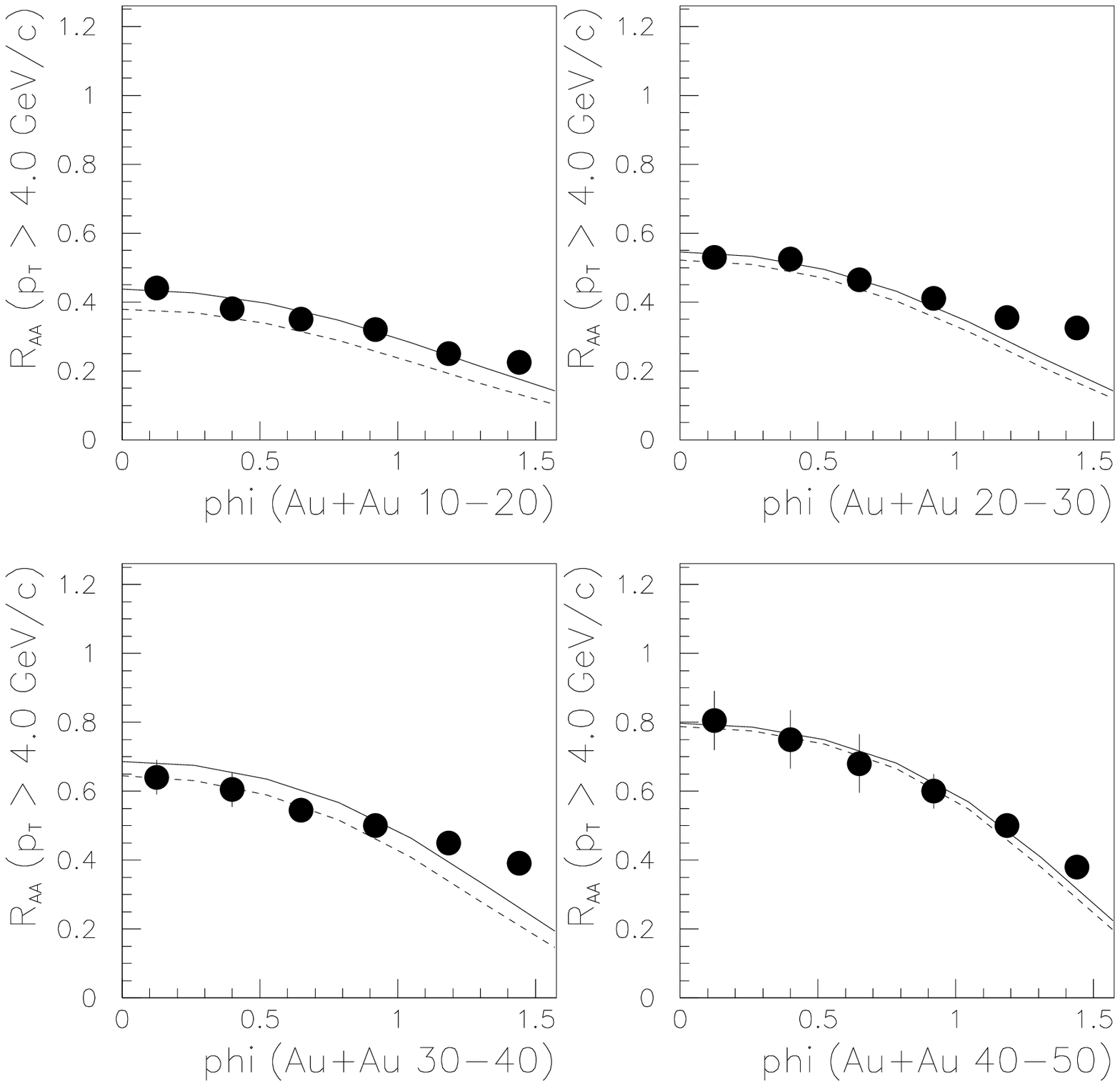,height=2.7in}
&
\psfig{figure=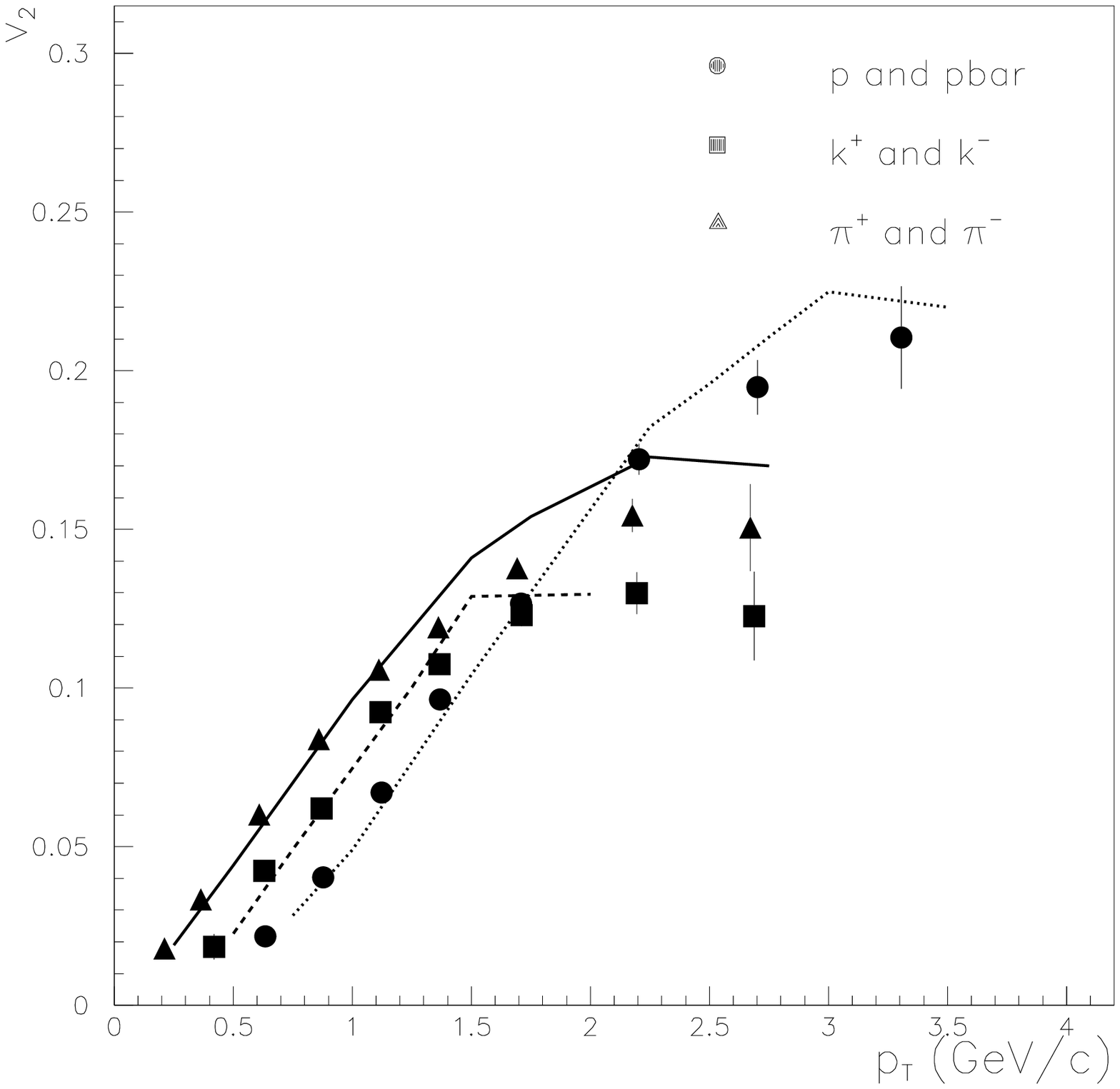,height=2.7in}\\
\end{tabular}
\end{flushleft}
\vskip -0.3cm
{\small Figure 3. Left: Values of 
$R_{AuAu}(b, \theta_R)$
as a function of the azimuthal angle $\theta_R$ for $p_T \geq 4$~GeV in various centrality bins.
Right: $v_2$ vs. $p_T$ of differents particles 
for mb collisions. Data \cite{3r} are from PHENIX}

\vskip 0.2cm
We have proposed a final state interaction model which takes into
account the different path length of a particle in the transverse
plane for each value of its azimuthal angle.
In our approach, the mechanism responsible for the large $p_T$ suppression
gives a contribution to $v_2$.
Elliptic flows very close to the experimental ones are obtained in the
whole $p_T$ region. 
Although this contribution to $v_{2}$ results from an asymmetry in the
azimuthal angle, it can be qualified as non-flow:
the mechanism from which it arises (fixed $p_T$ suppression) is maximal
at zero impact parameter and
thermalization is not needed.
We do not claim that 
our mechanism gives
the only contribution to the elliptic flow.
In our opinion, our knowledge of the dynamics of the nuclear
interaction is not sufficient 
to disentangle all the interpretations regarding the
$v_2$.


\end{document}